# Sequential Multi-Step Nanoimprint Lithography Fabrication of Zero-Mode Waveguide Nanoaperture Arrays to Enhance Single Molecule Fluorescence Detection

Hamza Khelidj,[1] Badre Kerzabi,[2] Igor Ozerov,[3] Olivier Hector,[1] David Grosso,[2,3] and Jérôme Wenger[1,*]

[1] *Aix Marseille Univ, CNRS, Centrale Med, Institut Fresnel, AMUTech, 13013 Marseille, France*

[2] *Solnil, 163 Avenue de Luminy, 13009 Marseille, France*

[3] *Aix-Marseille Univ, CNRS, CINaM, AMUTech, Campus de Luminy 13288, Marseille, France*

* *Corresponding author: jerome.wenger@fresnel.fr*

**Abstract:** Zero-mode waveguides (ZMWs) enable single-molecule fluorescence detection at micromolar concentrations by confining light to nanoscale volumes, overcoming the diffraction limit of confocal microscopy. However, their widespread adoption is hindered by high fabrication costs and limited throughput of traditional methods like focused ion beam or electron-beam lithography. Here, we introduce a scalable cost-effective approach using sequential nanoimprint lithography (NIL) combined with hydrofluoric acid etching to fabricate ZMW arrays with tunable diameters from a single initial master. In our sequential nanoimprint approach, each stamped NIL output serves as a master for the next nanoimprint generation. By leveraging the shrinkage of sol-gel nanopatterns during annealing, we achieve a cumulative diameter reduction from 230 nm to 115 nm over four successive imprints, all based on the same initial master. The resulting ZMWs exhibit detection volumes reduced by up to 1000-fold and fluorescence enhancement exceeding 16×, achieving performance comparable to state-of-the-art focused ion beam-fabricated devices. Eliminating the need for multiple master structures significantly expands the scalability of nanoimprint lithography approaches. By lowering the nanofabrication barriers and making ZMW arrays more accessible, the sequential NIL method paves the way towards broader adoption of nanophotonic devices in single-molecule biophysics, biosensing, and surface patterning applications.

**Figure for Table of Contents**

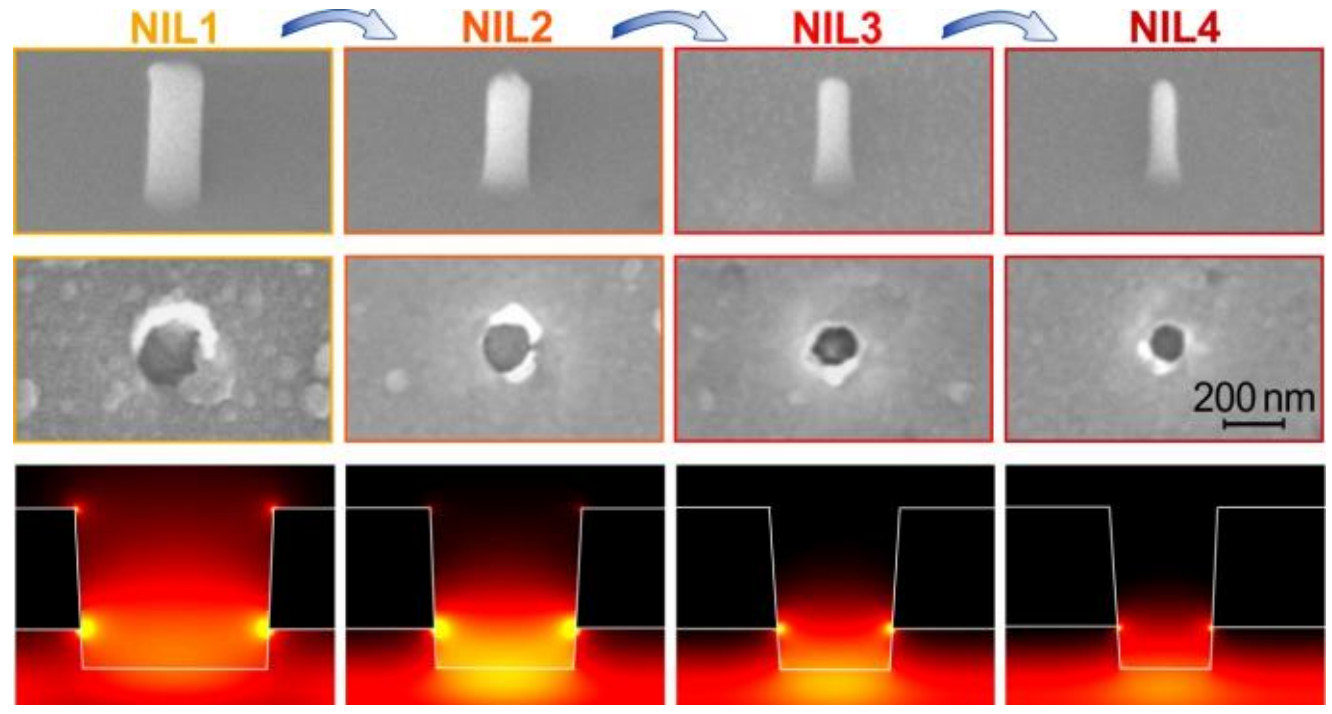

**Keywords:** zero-mode waveguide nanoaperture; nanoimprint lithography; nanofabrication; single molecule fluorescence; fluorescence enhancement

## Introduction

Zero-mode waveguide (ZMW) nanoapertures are subwavelength holes milled into an opaque metallic film, which have received a large interest to overcome the limitations of single-molecule fluorescence detection.[1] Although fluorescence techniques have become central tools to investigate molecular biophysics,[2–4] the optical detection of single molecules remains constrained by diffraction.[5,6] Due to the limited fluorescence brightness per emitter, the detection of single-molecule events is challenged and the temporal resolution is typically restricted to the millisecond range in order to collect enough photons per time bin.[7–9] A second limitation affects the necessary concentration range to isolate a single molecule: diffraction-limited microscopes with a typical confocal volume of 1 fL ($10^{-15}$ L) require concentrations in the pico to nanomolar range.[5,6] However, a significant range of biochemical reactions occur at micromolar to millimolar concentrations and thus cannot be probed at the single-molecule level on diffraction-limited microscopes.[10,11]

Both diffraction limitations affecting fluorescence brightness and detection volume can be simultaneously overcome by using ZMW nanoapertures.[12–15] In contrast to total internal reflection fluorescence (TIRF) microscopy, which constrains the detection volume along the longitudinal propagation axis, ZMWs confine light at the nanoscale along three dimensions.[16] This allows to isolate single molecules even inside micromolar concentrated solutions.[17–23] Thanks to the enhanced light-matter interaction the ZMW, the fluorescence brightness per molecule is improved, enabling to reach 10 µs temporal resolution or below.[9,24–28] ZMWs are relatively easy to use for experiments with molecules diffusing in solution, offering straightforward signal analysis. In contrast, plasmonic nanoantennas require careful two-species fitting to isolate the hot spot signal from the surrounding fluorescence background.[29–31]

While ZMWs bring many advantages to improve single-molecule fluorescence detection, their nanofabrication cost and limited availability remain major issues hindering their applications in biophysics [32–37] and sequencing of DNA and proteins.[38,39] Focused ion beam (FIB) and electron-beam lithography (E-beam) enable fine control of the samples,[40–46] yet at expensive purchase and maintenance costs that are often prohibitive. Nanosphere lithography (NSL) offers a reduced operational cost,[47–51] but imposes limitations on the ZMW shape and spacing.[41] To overcome these technological issues, we have recently introduced an alternative ZMW nanofabrication approach based on sol-gel nanoimprint lithography (NIL) combined with hydrofluoric acid (HF) vapor-phase etching.[52] NIL allows the replication of nanopillar arrays from a master structure at low cost, high throughput and high resolution.[53–57] The nanopillars are then selectively etched with HF vapor to mill the nanoapertures in a controlled and massively parallel fashion.[52]

While NIL offers an elegant solution to the ZMW nanofabrication challenge, its scalability and cost-efficiency remain limited by the need of a master structure often requiring E-beam lithography.

Changing the ZMW diameter necessitates a different master structure, adding to the fabrication cost and complexity to realize high-aspect-ratio pattern arrays. Our preliminary work was limited to ZMW nanoapertures of 230 nm diameter.[52] As a consequence of this relatively large diameter, the detection volume was only 100× below the confocal diffraction limit, and the enhancement factor of the fluorescence brightness was below 8×. Reaching higher performance requires smaller ZMW diameters, which in turn poses challenges to the fabrication and NIL replication of different master structures.

Here we introduce a sequence of multiple nanoimprint lithography steps to fabricate a series of ZMW samples with different diameters from the same initial master structure. Our approach takes advantage of an often overlooked feature of sol-gel nanoimprint: upon solvent evaporation and annealing, the size of the imprinted nanopattern decreases by 15 to 25%. The key concept here is to reuse the NIL-fabricated array as a master structure for the next replicate generation (Fig. 1). As a result of this sequential process, each NIL generation benefits from a cumulative reduction of the ZMW diameter, starting from 230 nm for the first generation (so-called NIL1) and going down to 115 nm for the fourth successive imprint (NIL4). The optical performance of the resulting ZMW arrays is characterized by fluorescence correlation spectroscopy (FCS), demonstrating detection volume reduction down to 1000× and fluorescence enhancement above 16×, significantly outperforming our earlier work,[52] and leading to results comparable to ZMWs fabricated by FIB.[9] Using a single master structure fabricated by E-beam to produce a range of structures with different sizes noticeably expands the scalability of NIL-based approaches. This sequential NIL fabrication strategy represents a significant step forward in making nanophotonic devices more accessible to single-molecule studies and other nanophotonic applications including biosensing,[58–60] filtering,[16] microscope calibration,[61] and surface patterning.[62]

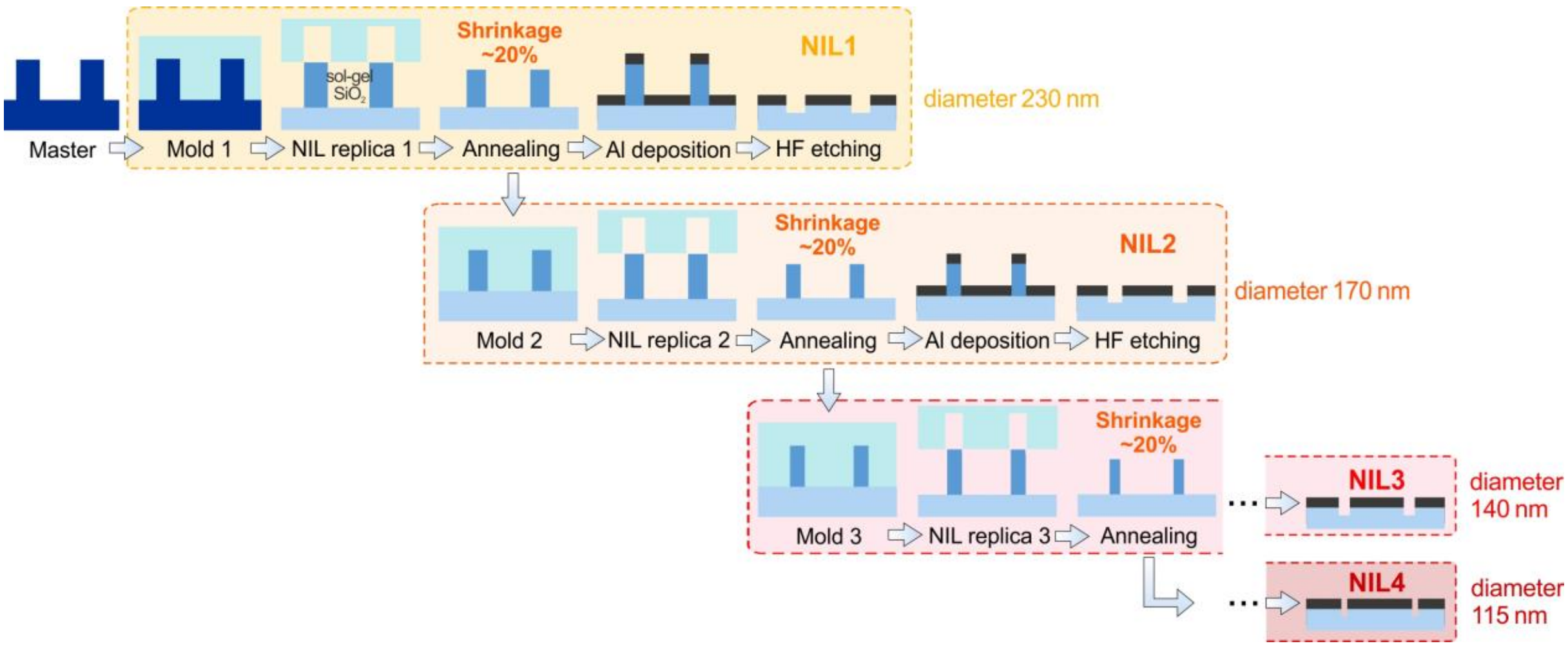

**Figure 1.** Sequential zero-mode waveguide nanoaperture fabrication methodology. Each generation step (color background surrounded by dashed line) merges sol-gel nano-imprint lithography with HF

vapor etching to produce ZMWs. Taking advantage of the sol-gel shrinkage during the NIL process, each replica can be further reused as a subsequent master structure with a reduced size. We iterate the process 4 times, yielding ZMWs structures named NIL*x*, where the number *x* indicates the amount of NIL replicas used.

## Results and Discussion

Figure 1 describes schematically our ZMW nanofabrication method based on multiple NIL steps followed by a final selective acid etching in vapor phase. The first step, denoted NIL1 in Fig. 1, is similar to our recent work described in Ref.[52] Briefly, the initial master is an array of silicon nanopillars made by E-beam lithography with a diameter of 270 nm, a height of 700 nm and a 3 µm spacing covering a 1 mm × 1 mm² area (See Methods Section for details). A perfluoropolyether (PFPE) mold is created from this master and is pressed as NIL stamp on a glass coverslip covered with a thin layer of sol-gel containing $SiO_2$ precursors.[55,56] After thermal curing and solvent evaporation, the nano-pillar array pattern is transferred on the coverslip and bears the physical properties of silica. A 150 nm thick opaque aluminum layer is then deposited on the nanopillar array. Aluminum is chosen for its low cost and a broad spectral range covering the entire visible region. The sample is then briefly immersed in an acid solution (Aluminium Etchant Type A) to dissolve the aluminum residues on the wall sides of the nanopillars. The ZMW apertures are finally created by etching simultaneously all the nanopillars using hydrofluoric acid etching in vapor phase. This process selectively dissolves the silica nanopillars without affecting the aluminum parts.[63,64] During the fabrication process, HF etching is used to create an undercut region approximately 50 nm into the glass substrate beneath the aluminum film. This undercut corresponds to the zone of highest intensity enhancement, resulting from constructive interference between the incident beam and the beam reflected by the zero-mode waveguide (ZMW) structure. The extension of this undercut volume can be controlled by adjusting the HF etching time, [52] enabling optimization of both the fluorescence brightness and the detection volume of the ZMW devices. Because of the fast HF etching times of a few seconds and the small volume of the ZMWs, the undercut volume exhibits variability from batch to batch, but remains nearly uniform within a given sample.

During the thermal annealing in the NIL process, the imprinted dry sol-gel intermediate material shrinks through condensation to reach the dense $SiO_2$ state.[52,55] The nanopillar dimensions are reduced to a diameter of 230 nm and a height of 580 nm, while the 3 µm spacing remains unchanged. The shrinkage arises from the evaporation of our water-ethanol solvent during annealing and is material-dependent. For our TEOS-based sol-gel solution, we observe an essentially isotropic shrinkage with average diameter reduction of 19% and a height reduction of 18% (see Supporting Information Tab.

S1). The consistency of these values across NIL1–NIL4 generations demonstrates the reproducibility of the process.

We take advantage this feature to fabricate a series of ZMW samples with different diameters from the same initial Ebeam master. The NIL-fabricated array for a given generation is reused as a master structure for the next generation in a sequential process (Fig. 1).[65,66] As a result, the different NIL generations benefit from a cumulative reduction of the apparent nanopillar diameter due to sol-gel shrinkage. We iterate this process 4 times, starting from 230 nm diameters for the first NIL generation (NIL1) and going down to 115 nm for the fourth sequential imprint (NIL4). Figure S1 of the Supplementary Information shows SEM images of the imprinted nanopillars. The remaining steps of the fabrication protocol (aluminum evaporation, aluminum etching on the nanopillar walls, HF etching of the $SiO_2$ nanopillars) are similar to the ones described for NIL1, only the HF etching time is adjusted according to the nanopillar dimensions (see details in the Materials and Methods section).

The resulting ZMW nanoapertures are characterized by SEM, with the main results being presented in Fig. 2. Arrays of ZMW nanoapertures featuring a quasi-circular profile are produced in a highly parallel fashion without requiring expensive Ebeam or FIB equipment. To complete the data in Fig. 2a,b, high magnification top-view and cross-cut SEM images of ZMWs are shown in the Supplementary Information Fig. S2 and S3 for the different NIL generations. Detailed information about the statistical dispersion of the ZMW diameters is presented in Fig. 2c with histograms and box-and-whisker plots. A clear reduction in the ZMW diameters is observed along the different NIL generations, in line with the reduction of the imprinted nanopillar dimensions. The relative error on the diameter (defined as the ratio of the standard deviation by the average value) is about 3% for the initial Ebeam master structure. While the NIL process increases this value to 6% for NIL1, the relative error stays within a range of 5 to 7% for the different subsequent NIL generations, indicating that the sequence of NIL imprints allow to maintain a good control on the sizes of the fabricated ZMWs (See Fig. S9a and further discussion about the data dispersion below).

Optical microscopy images of the different ZMW arrays are shown in Fig. 3a, covering a field of view of 105 x 80 $\mu m^2$. Some apertures are absent as a result of the NIL process (e.g., missing pillars on the master structure or imprint defects), yet these defects represent only 2–4% of the total ZMW available, with the defect ratio remaining nearly constant across the NIL1 to NIL4 generations. As the ZMW diameter is reduced, the transmission decreases and becomes more sensitive to small variations in the ZMW dimensions, shapes and potential aluminum residues deposited on the nanopillars. As an expected result, the transmission image shows more variations for NIL3 and NIL4 than the earlier generations. As the main purpose of these ZMW apertures is to improve the detection of fluorescent molecules, we prefer to focus on the fluorescence characterization of our samples rather than the optical transmission, which is a sensitive and indirect observation.[16]

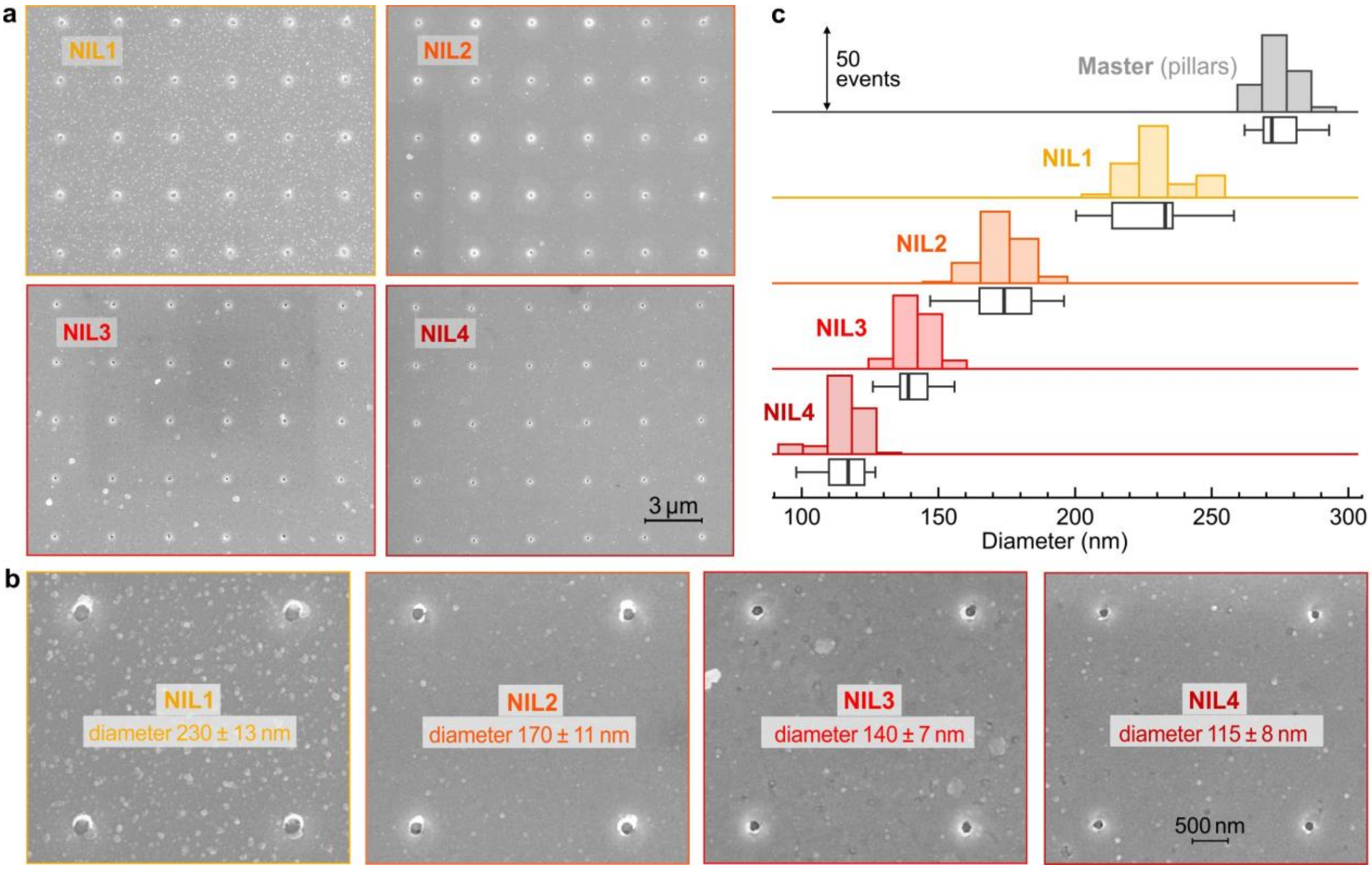

**Figure 2.** Zero-mode waveguide nanoaperture arrays with decreasing diameters fabricated by sequential NIL steps. (a,b) Scanning electron microscopy (SEM) images of the different samples. For each sample, the diameter of over 90 individuals ZMWs was measured from the SEM images. The values indicated in (b) correspond to the average ± one standard deviation. (c) Histogram of the diameters measured by SEM for the master structure and the subsequent NIL*x* structures. The box-and-whisker plots below each histogram indicate the median value (thick vertical line), the 25th and 75th percentiles (box dimensions) and the extrema of the distribution.

To assess the sample performance to improve fluorescence detection, we perform fluorescence correlation spectroscopy (FCS) experiments on Alexa Fluor 647 dyes diffusing in PBS solution.[52] The ZMW sample replaces the glass coverslip in a standard confocal fluorescence microscope operating at 633 nm, and is covered by the solution of Alex Fluor 647 molecules at micromolar concentration (see Methods section). Fluorescence correlation spectroscopy (FCS) is an established technique to measure the average number of molecules, their fluorescence brightness $CRM$ (count rate per molecule) and their average diffusion time.[67,68] FCS has also been extensively applied in the context of ZMWs.[9,13,35]

Raw FCS correlation data are shown in Fig. 3b for different solutions of calibrated concentrations, with supplementary FCS traces in Fig. S4 for a direct comparison between NIL samples.

Following the FCS analysis described in the Methods section, the number of observed molecules $N$ is recorded as a function of the dye concentration $c$ for the different NIL generations (Fig. 3c). We experimentally determine the ZMW detection volume $V_{eff}$ using a linear fit of the data in Fig. 3c and the relationship $c = N/(N_A V_{eff})$, where $N_A$ denotes Avogadro's number. Our results indicate that $V_{eff}$ scales from 10 aL ($10^{-18}$ L) for NIL1, and go down to below 1 aL for NIL4. As compared to a 1 fL confocal detection volume, these values represent a volume reduction ranging from 100 to 1000×, enabling FCS to be performed at micromolar concentrations.

The fluorescence brightness per molecule $CRM$ is a critical experimental parameter determining the precision and the temporal resolution of the measurements. The $CRM$ results for the different NIL samples are summarized on Fig. 3d as function of the excitation power, enabling to distinguish between the linear and the saturated regimes. The highest fluorescence enhancement and brightness are obtained using the ZWMs with 140 nm diameter corresponding to NIL3 with a gain of 16.6× compared to the confocal reference. This result agrees with prior data obtained on ZMWs fabricated by FIB,[9,24,69] and is a consequence of the trade-off between local excitation and emission enhancement versus the non-radiative losses into the metal.[70]

In the case of aluminum ZMWs, the fluorescence brightness enhancement is higher in the linear regime, taking advantage of cumulative gains in the collection efficiency, quantum yield and excitation rate (Fig. S5).[24,52] In the saturated regime, the enhancement no longer benefits from the higher excitation intensity inside the ZMW, but reveals the maximum count rate achievable per molecule, here above 1 million counts/s for NIL3. While the presence of the reflective aluminum surface increases the background intensity level as compared to the confocal reference (Fig. S6), the detected brightness per molecule largely exceeds this residual background. Photothermal heating effects are mitigated by the metal layer acting as a heat sink and were previously shown to play a minor role in similar ZMW structures.[9] This allows us to rule out photothermal heating as the cause of the observed plateau in Fig. 3d.

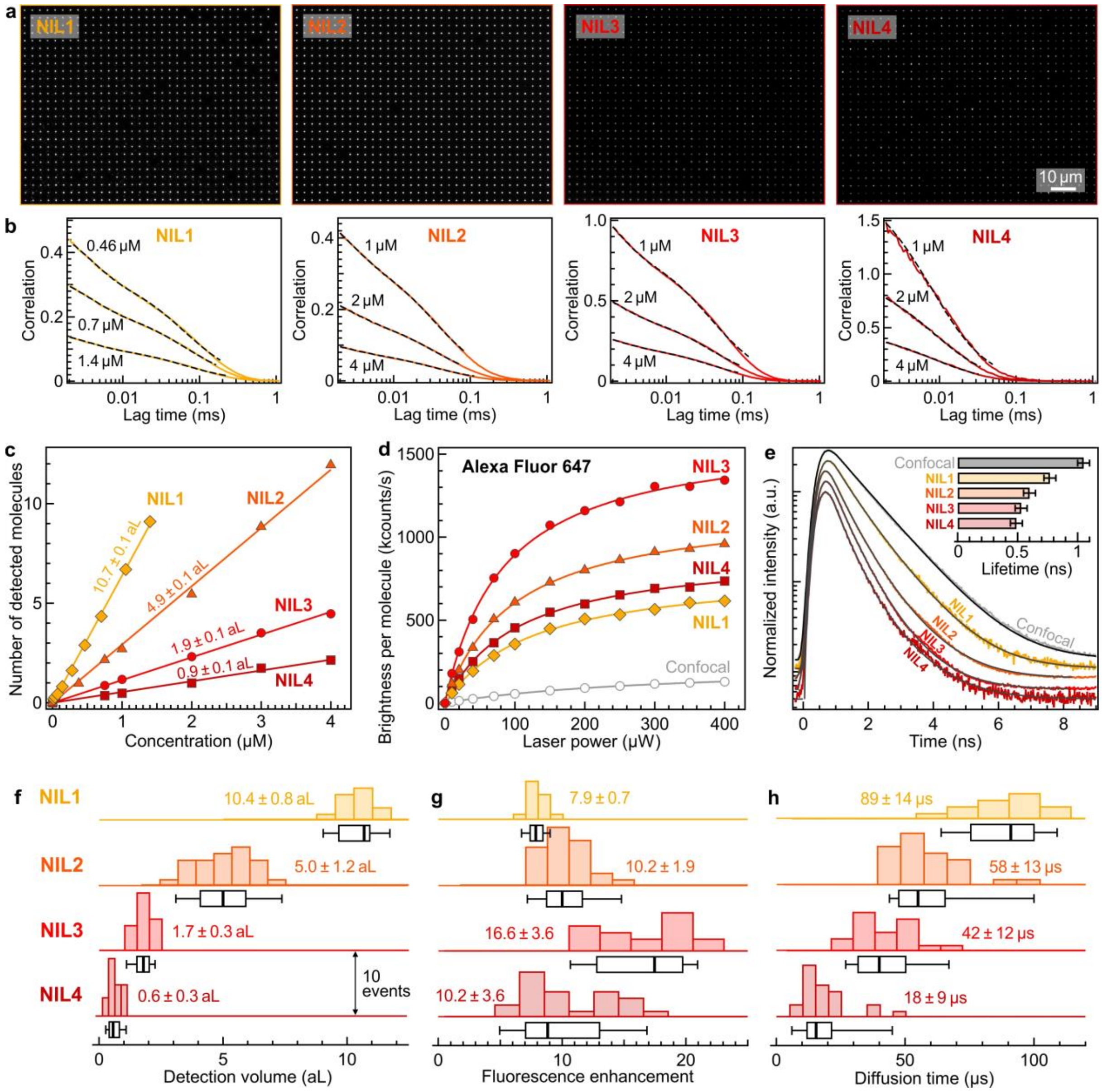

**Figure 3.** Characterization of ZMW apertures to enhance single molecule fluorescence detection. (a) Optical image of the ZMW array recorded with a 40x 0.6NA air objective. Each image covers a field of view of 105 x 80 µm², but the ZMW array is much larger with dimensions around 1 x 1 mm².[52] (b) FCS correlation data for diffusing Alexa Fluor 647 molecules at different concentrations. Dashed lines are numerical fits. (c) Number of molecules $N$ detected by FCS as function of the dye concentration for the different samples. Lines are numerical fits used to determine the effective detection volume. The error bars correspond to uncertainty resulting from the data interpolation. (d) Fluorescence brightness per molecule $CRM$ as function of the laser excitation power showing a clear enhancement as compared to the confocal reference. (e) Normalized fluorescence lifetime decays corresponding to the different NIL*x* samples. The traces are vertically shifted for clarity. Black lines are numerical fits. (f-h) Histograms of the ZMW detection volume (f), fluorescence enhancement of the brightness per molecule at 10 µW excitation power (g) and FCS diffusion time (h). The box-and-whisker plots below each histogram

indicate the median value (thick vertical line), the 25$^{th}$ and 75$^{th}$ percentiles (box dimensions) and the extrema of the distribution. The values written next to each histogram correspond to the average ± one standard deviation.

The supplementary Fig. S7 shows the evolution of the number of detected molecules and their diffusion times as function of the excitation power. For NIL1, we observe an increase in the number of detected molecules along with a nearly constant FCS diffusion time up to 200 µW. These are the typical signatures of fluorescence saturation in FCS at elevated excitation power.[67] For the NIL2 and NIL3 samples, increasing the laser power above 50 µW reduces the number of detected molecules, accompanied by a reduction in diffusion time (Fig. S7). These observations indicate the occurrence of photobleaching as the dyes diffuse across the ZMW volume. The effect is more pronounced for NIL3 which exhibits the highest fluorescence enhancement. Interestingly, in the case of NIL4, the number of detected molecules and the FCS diffusion time remain nearly constant up to 300 µW. We attribute this effect to the lower excitation intensity gain and the shorter lifetime, which partly mitigate the photobleaching effects in the case of NIL4.

To support the fluorescence enhancement observation, we have measured lifetime decay traces corresponding to our different NIL samples. The results summarized in Fig. 3e indicate a gradual reduction of the lifetime as the ZMW diameter is reduced, from 1 ns in the confocal reference to 0.5 ns in the 115 nm NIL4 structure. These findings are consistent with the acceleration of the photokinetic rates generally observed for ZMWs,[24,70,71] and confirm the validity of the ZMW performance assessment. As the fluorescence lifetimes for NIL3 and NIL4 remain quite comparable, we can conclude that the lower brightness found for NIL4 as compared to NIL3 is rather due to a decrease in the different gains (excitation, quantum yield and collection efficiency) rather that the occurrence of strong quenching for the 115 nm diameter of NIL4. The numerical simulations shown in Fig. S10 further support this conclusion.

The statistical distributions of the key measurement outputs are summarized in Fig. 3f-h for the detection volume, the fluorescence brightness enhancement and the FCS diffusion time. It is important to stress that these distributions include the variations of the ZMW dimensions due to fabrication imperfections together with all the experimental noise and uncertainties related to FCS measurements (positioning respective to the laser focus, shot noise, electronic noise, fit uncertainties…). Scatters plots of the diffusion time versus the detection volume and brightness enhancement versus the detection volume are shown in the Supplementary Fig. S8. The reduction in the ZMW diameter yields to the expected trend of decreased detection volume and faster diffusion times. For the fluorescence enhancement, we find that a diameter of about 140 nm (corresponding to a volume around 2 aL) yields the highest value, as a trade-off between local enhancement and coupling losses similar to earlier results achieved with FIB-based ZMWs,[24,26,69] or derived from numerical simulations.[70] Although numerical simulations indicate that NIL2 exhibits the highest local excitation intensity, the smaller diameter of NIL3 improves optical confinement, resulting in superior spatially-averaged fluorescence performance, consistent with numerical predictions.[26,70] In contrast, due to its smaller diameter, NIL4 suffers from both increased non-radiative losses,[24] and reduced excitation intensity gain (Fig. S10), explaining its lower brightness despite a shorter lifetime.

The relative dispersion of the samples (corresponding to the ratio of the average value by the standard deviation) increases with the NIL steps, rising from approximately 15% for NIL1 to about 45% for NIL4 (Fig. S9). This trend reflects the growing statistical dispersion in diameter as each successive NIL step introduces additional variability, primarily due to cumulative uncertainties in sol-gel shrinkage. Since these sources of noise are uncorrelated, the total variance scales additively, resulting in a standard deviation that increases by a square root factor. Consequently, the uncertainty in diameter remains relatively modest, reaching 7% for NIL4 as compared to 5–6% for NIL1–NIL3. The larger statistical dispersion observed in the FCS data for NIL4 arises from the diameter dependence of the ZMW performance. As the diameter decreases, light confinement within the aperture intensifies, sharpening the axial intensity decay and increasing optical losses. These effects amplify the sensitivity of ZMW performance to diameter variations. For NIL3, which yields the best overall FCS performance, the relative errors are approximately 20% for both the enhancement factor and detection volume, and 28% for the diffusion time. If needed, narrower statistical dispersions could be achieved by selecting and calibrating a subset of ZMW apertures before use for biophysical applications. Besides, we anticipate that further optimization of the different fabrication steps could help achieve an improved precision.

Numerical simulations of the illumination profile and detection volume corroborate our experimental findings, as shown in Fig. 4 and S10. The undercut region milled approximately 50 nm into the glass substrate beneath the aluminum film corresponds to the zone with the highest intensity enhancement (Fig. 4a). This phenomenon arises from constructive interferences between the incident beam and the beam reflected by the ZMW structure.[26,70,72] The contribution from this undercut region is therefore an essential element defining the ZMW performance. Figure 4b presents the simulations of the FCS detection volume using the approach described in refs.[73,74] The experimental observations align with the numerical predictions, though uncertainty in the undercut depth introduces some margin when interpolating the experimental data. Notably, the distributions of FCS observation volumes in Fig. 3f encompass all sources of uncertainty, including variations in ZMW diameter and undercut depth, shape variations due to metal grains, positioning of the ZMW relative to the laser focus, and FCS fit analysis. Although undercut depth may slightly vary along the ZMW array, we currently cannot isolate this contribution from other sources of uncertainty. The simulations in Fig. 4b further indicate that variations in ZMW diameter exert a major influence on the observation volume spread. We believe the undercut depth primarily acts as a systematic offset parameter for tuning the observation volume, but we cannot exclude that its variations contribute to the statistical distributions in FCS results. Additionally, the simulations in Fig. 4b indicate how much the detection volume can be tuned by changing the undercut depth for a specific ZMW diameter. Beyond milling all the ZMWs apertures simultaneously, the HF etching technique offers some control over the undercut depth into the glass substrate by adjusting the etching time, providing an additional advantage for tailoring ZMW properties.[52]

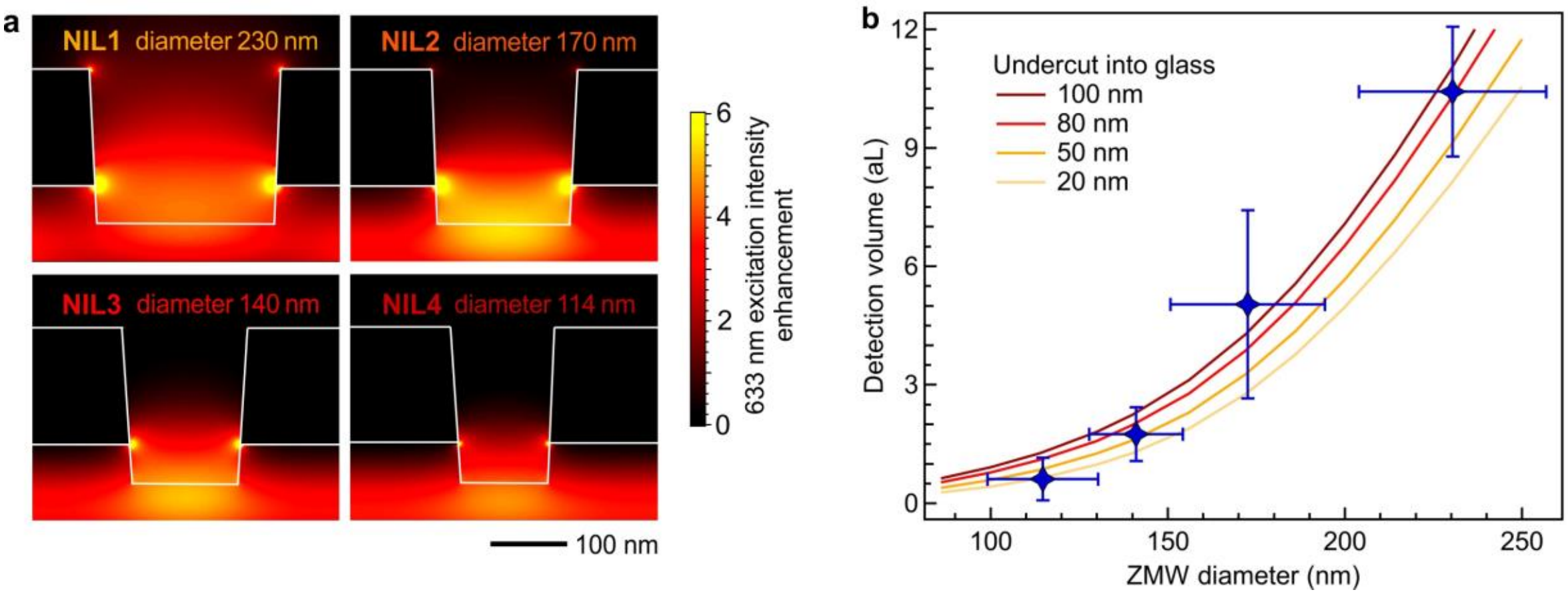

**Figure 4.** (a) Finite-difference time-domain (FDTD) simulations of the excitation intensity inside the ZMWs of different diameters. The ZMW design includes a 50 nm undercut volume extending into the glass substrate. (b) Detection volume computed using the formula $V_{eff} = (\int I_{ex}\, MDE\; dV)^2 / \int (I_{ex}\, MDE)^2\; dV$ as a function of the ZMW diameter.[73,74] The integration extends to different values of the undercut into the glass substrate as noted on the graph. The blue crosses indicate the experimental average data, with the error bars corresponding to 2× the standard deviation to better represent the experimental distributions in Fig. 2c and 3f.

## Conclusions

We have developed a scalable cost-effective methodology to fabricate ZMW arrays with a range of different diameters from a single initial master structure. The central innovation lies in the use of sequential nanoimprint lithography steps taking advantage of the shrinkage phenomenon occurring during sol-gel annealing. NIL-fabricated arrays serve as master pattern for the next generation, enabling the production of ZMW samples with progressively reduced diameters. This cumulative reduction in the dimensions importantly avoids the need for the expensive nanofabrication of different master structures. Through this iterative process, diameters decrease from 230 nm in the first-generation NIL1 to 115 nm in the fourth-generation NIL4. The high optical performance of the resulting ZMW arrays is demonstrated by FCS characterization, demonstrating reduction of the detection volume down to 1000× and fluorescence enhancement above 16×, leading to results comparable to ZMWs fabricated by FIB.[9,69]

The sequential NIL process can be extended to a fifth generation (NIL5), and remains effective despite the small diameter of the NIL4 master (Fig. S11). However, the resulting nanopillars exhibit a distorted conical shape with an enlarged base ~120 nm, yielding a ZMW diameter similar to NIL4. This shape distortion currently sets the limits for the number of sequential imprint generations.

While NIL and HF etching are individually established in microelectronics and microelectromechanical systems, their combination opens a promising solution for large scale fabrication of nanophotonic devices at moderate operational costs. The method's capacity for precise cumulative reduction in device dimensions eliminates the need for multiple master structures and significantly expands the scalability of NIL-based approaches. By lowering the technological barriers to ZMW fabrication, this work contributes to a broader adoption of nanophotonic tools in single-molecule biophysics, overcoming the limitations of confocal and TIRF-based detection schemes. The improved accessibility to nanoaperture arrays also benefits to related fields, including biosensing,[58–60] optical filters,[16] microscope calibration,[61] and surface patterning.[62]

## Materials and Methods

**ZMW samples fabrication.** The ZMW nanoapertures were fabricated using a combination of nanoimprint lithography (NIL) and vapor-phase selective etching (VPE). The progressive reduction in diameter of the structures was achieved through a shrinkage phenomenon that occurred after each NIL step. The different steps of the sequential NIL process are detailed below.

The initial silicon master was fabricated using electron beam lithography (EBL) with a Raith Pioneer system, followed by reactive ion etching in an $SF_6$-$O_2$ plasma. The nanopillar array covered an area of 1 mm × 1 mm² on a monocrystalline silicon (001) substrate. Each pillar had a diameter of 270 nm, a height of 700 nm, and a periodic spacing of 3 µm.

Each individual NIL step followed the approach introduced in Ref.[52] Molds made of perfluoropolyether (PFPE) were fabricated from the selected structure. The pattern was transferred onto a Si substrate by NIL using a sol-gel solution containing tetraethyl orthosilicate (TEOS) as a $SiO_2$ precursor and a water-ethanol mixture as solvent (Solnil solution C). Before nanoimprint, the sol–gel layer was deposited by spin coating at 4000 rpm for 10 s, followed by the application of the mold. The mold-substrate assembly was heated to 70 °C for 2 minutes to promote solvent evaporation and pattern stabilization. After demolding, the structures were thermally annealed at 500 °C for 30 minutes. Starting from the Ebeam master structure, the first step, referred to as NIL1, produced nanopillars with a diameter of 230 nm and a height of 580 nm, corresponding to about 15% reduction in diameter due to shrinkage.

The nanopillars obtained after NIL1 were used as a new master for fabricating a PFPE mold. The pattern was subsequently transferred onto a Si substrate by NIL following the same protocol as described previously. This step, denoted NIL2, yielded nanopillars with a diameter of 170 nm and a height of 480 nm, corresponding to an additional diameter reduction of approximately 26%. The characteristic dimensions of the nanopillars from different NIL generations are summarized in the Supporting Information Tab. S1.

After the NIL1 and NIL2 steps, a decrease in the aspect ratio was observed due to shrinkage. To restore an aspect ratio suitable for subsequent processing, a reactive ion etching (RIE) step was performed for 105 seconds using a Plassys MG 200 system. The process used gas flow rates of 20 sccm of $SF_6$ and 8 sccm of $O_2$, under a pressure of 22 mTorr and a power of 120 W provided by a 13.56 MHz RF generator. Following this etching, the pattern was transferred onto Si substrate by a new NIL step according to the previous protocol. This third step, denoted NIL3, led to the formation of nanopillars with a diameter of 140 nm and a height of 460 nm, corresponding to a further reduction of approximately 18% in diameter.

The nanopillars from NIL3 were used as a master to fabricate a new mold in PFPE. The pattern was then again transferred onto Si substrate by NIL under the same experimental conditions. This fourth step, noted NIL4, led to nanopillars with a diameter of 115 nm and a height of 390 nm, representing a supplementary reduction of approximately 18% in diameter.

The patterns obtained from NIL1 to NIL4 were subsequently transferred onto cleaned borosilicate glass coverslips (1 inch diameter, 150 µm thickness) by NIL following our previously described protocol.[52] A RIE step was carried out for 1 minute to remove the residual $SiO_2$ layer between the nanopillars, thereby ensuring selective etching in the subsequent steps, using the same system and parameters as before. A 100 nm Al thin film was deposited at room temperature by electron-beam physical vapor deposition (EBPVD, Bühler Syrus Pro 710) using a high-purity (99.99%) Al target and a deposition rate of 10 $nm.s^{-1}$.The deposition chamber was maintained under vacuum at approximately $1 \times 10^{-7}$ mbar, and the substrate holder was rotated at 30 rpm to ensure a uniform film thickness. The residual aluminum deposited around the nanopillars was then partially removed by wet etching, using "Aluminum Etchant Type A" (Sigma Aldrich), a mixture of phosphoric acid (>77%), nitric acid (<3%), and acetic acid (<3.5%), before rinsing with MilliQ water and drying with nitrogen. The Al etching times were 40 s for NIL1 and 10 s for NIL2-NIL4. The selective etching of silica nanopillars was then carried out by HF vapor-phase etching using a VPE100 system (Idonus), with an evaporation temperature maintained at 35°C using a 10% liquid HF solution. The HF etching times were 50 s for NIL1, 40 s for NIL2, 34 s for NIL3 and 28 s for NIL4. Following the HF etching process, the samples were submerged in MilliQ water and exposed to an ultrasonic bath for 20 seconds in order to eliminate etching residues. To ensure a good wettability of the Al and glass surfaces before the FCS experiments, the ZMW arrays were cleaned for 4 min with a plasma cleaner (Diener Femto 50W, 0.6 mbar air).

Scanning electron microscopy (SEM) images were recorded with an FEI DualBeam DB235 Strata microscope equipped with a dual beam of gallium ions used to realize cross-sections of the nanoapertures at 30 kV and 100 pA.

The silicon master and intermediate silica molds have been reused multiple times (15 and 5 uses, respectively) without degradation, while PFPE molds reliably produce ~10 ZMW chips before replacement, confirming the robustness of our fabrication approach. The minor shape and side-wall roughness variations present are primarily due to metal grains and remaining aluminum, rather than the sequential NIL process itself, indicating preserved replication fidelity across generations, see Fig. S12 for a comparison of NIL1 and NIL4 SEM images.

**Fluorescent molecule samples.** Carboxylate-modified Alexa Fluor 647 dyes are purchased from ThermoFisher and used as received. All experiments are performed in standard 1x Phosphate-Buffered Saline (PBS) solution containing 140 mM NaCl, 10 mM phosphate buffer, and 3 mM KCl, pH 7.4 at 25°C.

**Optical microscope***.* All the optical characterization experiments are performed using a custom-built confocal microscope (Nikon Ti-U Eclipse) similar to earlier studies.[9,52] We use a pulsed laser diode at 635 nm (Picoquant LDH, 70 ps duration, 40 MHz repetition rate) and a helium-neon continuous laser at 633 nm (Melles Griot) for the power dependence study in Fig. 3d. All the laser beams pass through the same polarization-maintaining single mode optical fiber (Thorlabs) defining a similar beam profile at the microscope entrance. The laser polarization is linear and the power is 10 µW if not stated otherwise. A multiband dichroic mirror (ZT 405/488/561/640rpc, Chroma) reflects the laser beam towards the microscope. The molecules are excited using a 63x, 1.2 NA water immersion (Zeiss C-Apochromat) and the emission is collected by the same objective in epifluorescence configuration without polarization selection. To reject the backscattered laser light, the emission is passed through the same multiband dichroic mirror and a multi-band emission filter (ZET405/488/565/640mv2, Chroma). 50 µm confocal pinhole is conjugated to the sample plane. The detection channel uses a SPCM AQR 13 avalanche photodiode (Perkin Elmer). The detection in the 650-750 nm spectral region is performed with a ET655LP filter (Chroma). The photodiode signal is connected to a HydraHarp400 single photon counting module (Picoquant) in a time-tagged time-resolved (TTTR) mode. Fluorescence lifetime measurements have a temporal resolution of 400 ps upon 635 nm excitation, mainly limited by the APD jitter.

**FCS analysis.** All the analysis is performed using Symphotime64 software (Picoquant). The FCS curves are fitted using a three dimensional Brownian diffusion model with an additional term to include the blinking effects.[67,68]

$$G(\tau) = \rho\left[1 + \frac{T_{ds}}{1 - T_{ds}}\exp\left(-\frac{\tau}{\tau_{\mathrm{ds}}}\right)\right]\left(1 + \frac{\tau}{\tau_{\mathrm{d}}}\right)^{-1}\left(1 + \frac{1}{s^2}\frac{\tau}{\tau_{\mathrm{d}}}\right)^{-0.5} \quad (1)$$

Where $\rho$ is the correlation amplitude, $T_{ds}$ is the fraction of molecules going in the dark state, $\tau_{\mathrm{ds}}$ is the blinking time of the dark state, $\tau_{\mathrm{d}}$ is the mean diffusion time, and $s$ corresponds to the aspect ratio of the axial to the transversal dimension of the excitation source in the solution. For the confocal case, we used a value of $s$=5, whereas for the ZMW we used a value of $s$=1 based on our past results.[9,24,52,69] From the correlation amplitude $\rho$, the measured background $B$ on the ZMW (Fig. S5), and the total detected intensity $F$, we extract the average number of molecules $N$ and the brightness or count rate per molecule $CRM$ as:[9]

$$N = \frac{1}{\rho}\left(1 - \frac{B}{F}\right)^2 \quad (2)$$

$$CRM = \frac{1}{N}(F - B) \quad (3)$$

$N$ corresponds to the average number of detected fluorescent molecules in the observation volume over the duration of the experiment.[67] The evolution of the brightness per molecule $CRM$ as a function of the excitation power P is given by [9,75]

$$CRM = \mathrm{A}\frac{\mathrm{P}}{1+\frac{\mathrm{P}}{\mathrm{P_{sat}}}} \quad (4)$$

where $\mathrm{A} = \eta\phi\sigma\rho$ is the product of collection efficiency $\eta$, the quantum yield $\phi$, the excitation cross-section $\sigma$ and a constant proportionality parameter $\rho$ to accommodate for the different units. $\mathrm{P_{sat}}$ represents the saturation power.

**Fluorescence lifetime analysis.** The fluorescence lifetime decay histograms are fitted with a bi-exponential function reconvoluted by the instrument response function (IRF). All the analysis is performed using Symphotime64 software (Picoquant). The long lifetime component is kept constant at 1 ns corresponding to the lifetime of Alexa Fluor 6476 dyes in a free open solution. This contribution typically amounts for 20 to 30% of the total photons. The short lifetime component is left as fit parameter. As the local density of optical states inside a ZMW depends on the position and orientation of the emitting dipole, our data will necessarily result from a spatial and orientation averaging. The indicated lifetime corresponds to the amplitude-averaged lifetime between the two components.

**Numerical simulations.** The electric field intensity maps shown in Fig. S10 are computed with finite-difference time-domain (FDTD) method using RSoft Fullwave software. The ZMW geometry is set according to the experimental parameters deduced from the SEM images. The undercut depth into the glass substrate is set to 50 nm. The aluminum complex refractive index is taken from ref. [76]. The borosilicate glass substrate has a refractive index of 1.52, with the ZMW inner volume and top space being filled with water (refractive index 1.33). The simulations use 1 nm mesh size along the transverse dimension and 2 nm along the longitudinal axis. Over 8 optical periods are computed to ensure convergence. The numerical simulations of the detection volume in Fig. 4b use the formula $V_{eff} = (\int I_{ex}\, MDE\; dV)^2 / \int (I_{ex}\, MDE)^2\; dV$ following the approach developed in Refs.[73,74] with the excitation intensity profiles $I_{ex}$ and the molecular detection efficiency $MDE$ shown in Fig. S10.

**Statistical Analysis.** No pre-processing nor filtering was applied for the FCS experiments. The data presentation correspond to the average value with the error bars corresponding to the standard deviation. The statistical analysis was performed using Symphotime64 (Picoquant) and IgorPro 7 (Wavemetrics).

**Supporting Information**

SEM images of nanoimprinted pillar replicas; Top view SEM images of ZMW apertures; Cross-cut views of ZMW apertures; Comparison of FCS correlation data for different ZMW diameters; Background intensity as function of laser power; Power dependence of the number of molecules and diffusion times; Comparison of the enhancement factors in the linear and saturated regimes; Scatter plots of detection volume, diffusion time and fluorescence enhancement; Evolution of the relative errors across NILx samples; Numerical simulations; Fifth-generation nanoimprint; Comparison of NIL1 and NIL4 SEM images; Size reduction across NIL generations.

**Acknowledgments**

The authors thank Anthony Gourdin, Mohammed Bouabdellaoui, and Antonin Moreau for stimulating discussions. Nanofabrication was carried out using the PLANETE and Espace Photonique technological platforms, which are part of the CT PACA, and RENATECH+ French national technological network.

**Funding Sources**

This project has received funding from the European Research Council (ERC) under the European Union's Horizon 2020 research and innovation programme (grant agreement No 101122563), from the Centre National de la Recherche Scientifique (CNRS) under the Prematuration programme, and from Aix Marseille Université under the AMUTech programme.

**Conflict of Interest**

B.K. is employee of the Solnil company. D.G. is a co-founder of the Solnil company. The remaining authors declare no conflict of interest.

**Data Availability Statement**

The data that support the findings of this study data are available from the corresponding author upon reasonable request.

**Table of Contents Entry** For Table of Contents Use Only

**Sequential Multi-Step Nanoimprint Lithography Fabrication of Zero-Mode Waveguide Nanoaperture Arrays to Enhance Single Molecule Fluorescence Detection**

Hamza Khelidj, Badre Kerzabi, Igor Ozerov, Olivier Hector, David Grosso, and Jérôme Wenger

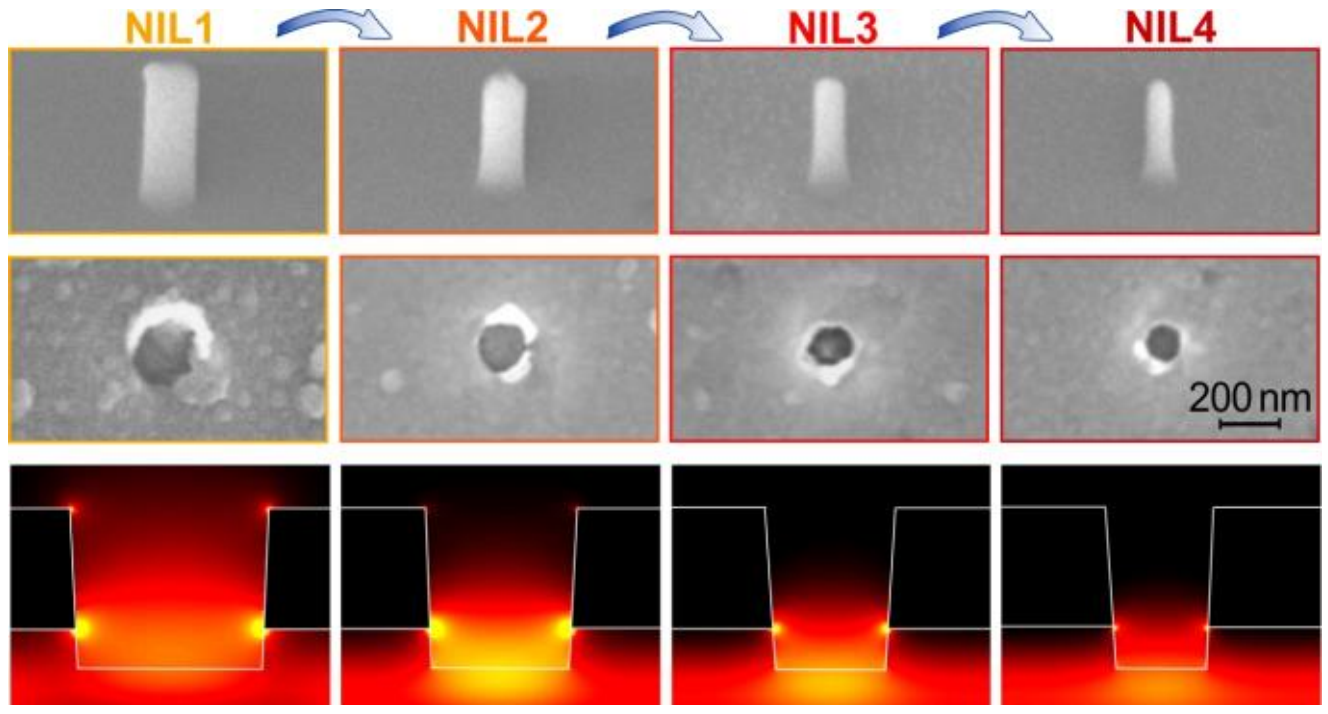

**Short synopsis:**

A scalable strategy is presented to fabricate arrays of zero-mode waveguide (ZMW) nanoapertures based on sequential nanoimprint lithography (NIL). Each NIL-generated replica is reused as a master, taking advantage of exploiting the intrinsic shrinkage of sol–gel materials during annealing to enable a cumulative dimension reduction from 230 nm down to 115 nm. This approach circumvents the need for multiple high-resolution master structures and significantly enhances fabrication scalability, enabling a broader accessibility of nanophotonics for single-molecule fluorescence applications.

## Supporting Information for

# Sequential Multi-Step Nanoimprint Lithography Fabrication of Zero-Mode Waveguide Nanoaperture Arrays to Enhance Single Molecule Fluorescence Detection

Hamza Khelidj,[1] Badre Kerzabi,[2] Igor Ozerov,[3] Olivier Hector,[1] David Grosso,[2,3] and Jérôme Wenger[1,*]

[1] *Aix Marseille Univ, CNRS, Centrale Med, Institut Fresnel, AMUTech, 13013 Marseille, France*

[2] *Solnil, 163 Avenue de Luminy, 13009 Marseille, France*

[3] *Aix-Marseille Univ, CNRS, CINaM, AMUTech, Campus de Luminy 13288, Marseille, France*

* *Corresponding author: jerome.wenger@fresnel.fr*

**Contents:**

## S1. Nanoimprinted pillar replicas

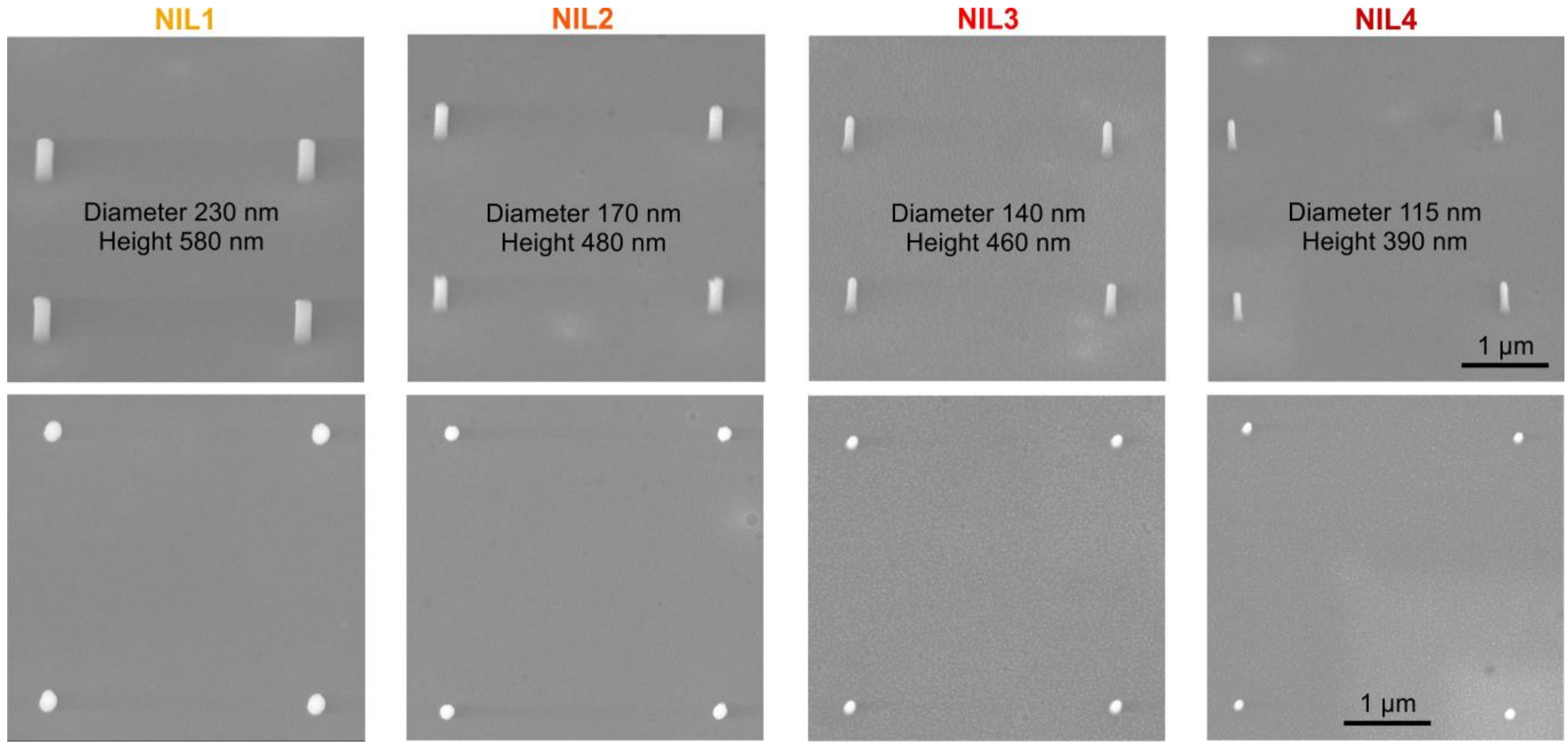

**Figure S1.** SEM images of the nanopillars fabricated using sequential NIL process.

## S2. Top view SEM images

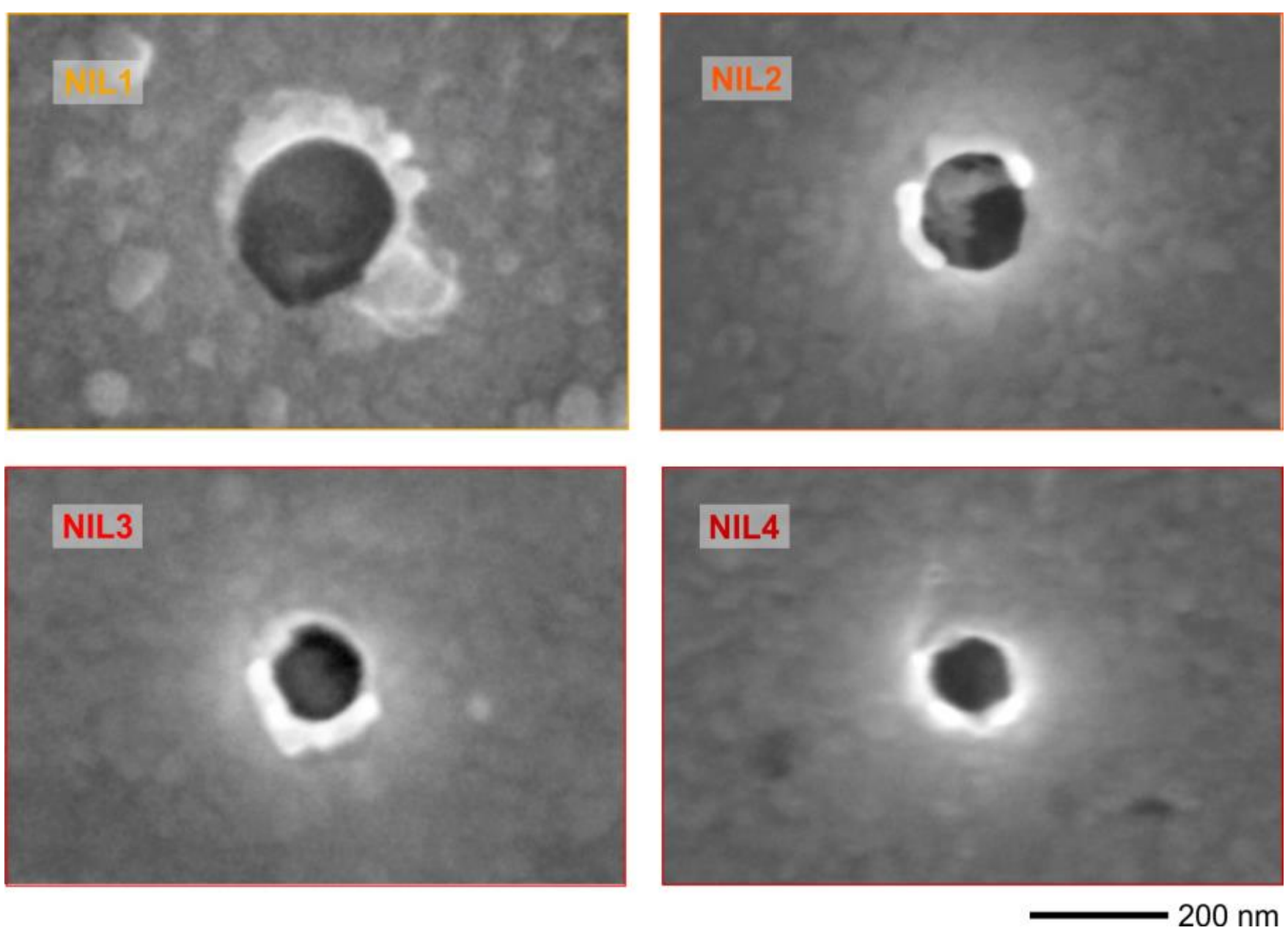

**Figure S2.** Top view SEM images corresponding to the sequential NIL*x* fabrication. For some samples, such as the NIL2 shown here, some residues of the silica pillar remain in the ZMW. While this contributes to enlarge the statistical dispersion of our results, on average the values still converge towards reduced detection volume and enhanced fluorescence brightness per molecule (Fig.3h,g).

## S3. Cross-cut views of ZMW nanoapertures

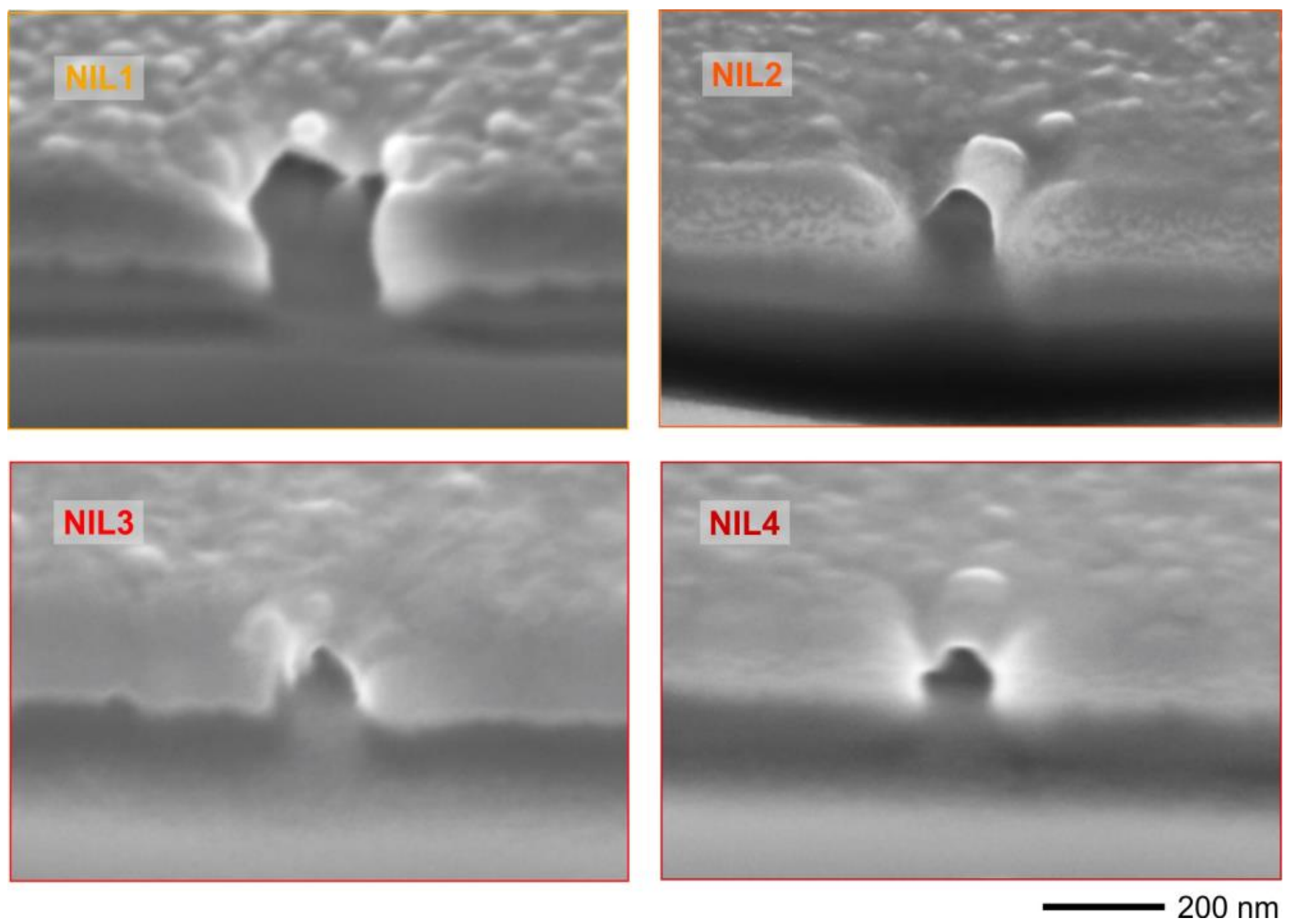

**Figure S3.** Cross-cut views of single ZMW apertures fabricated using the approach in Fig. 1. Approximately half of the sample has been cut by focused ion beam (FIB) milling. The sample surface is tilted by 52° respective to the SEM imaging direction.

## S4. Comparison of FCS correlation data for different ZMW diameters

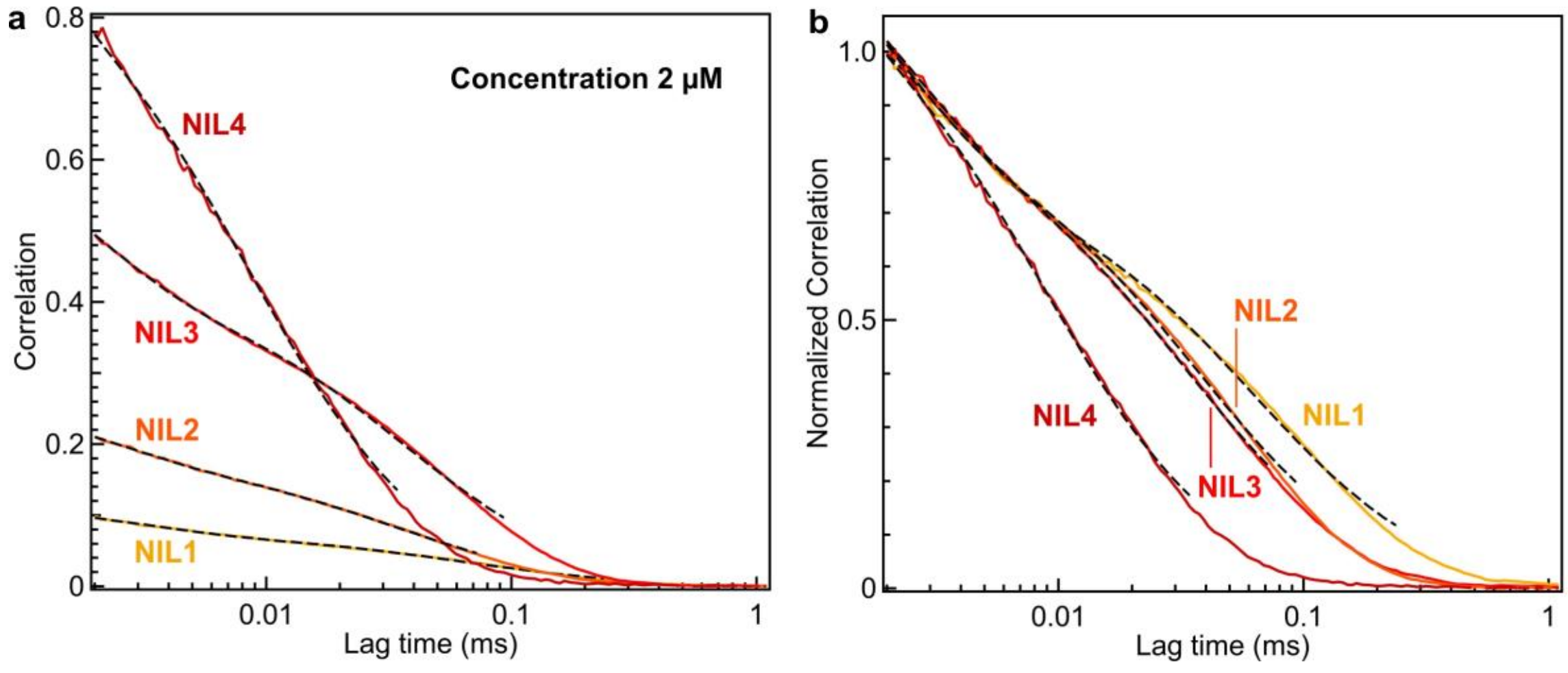

**Figure S4.** (a) FCS correlation data (solid color lines) and numerical fits (dashed lines) for the different NIL samples at 2 µM Alexa Fluor 647 concentration. Larger volumes lead to reduced correlation amplitudes. (b) Normalized traces in (a) to better show the differences in correlation times.

## S5. Comparison of the enhancement factors in the linear and saturated regimes

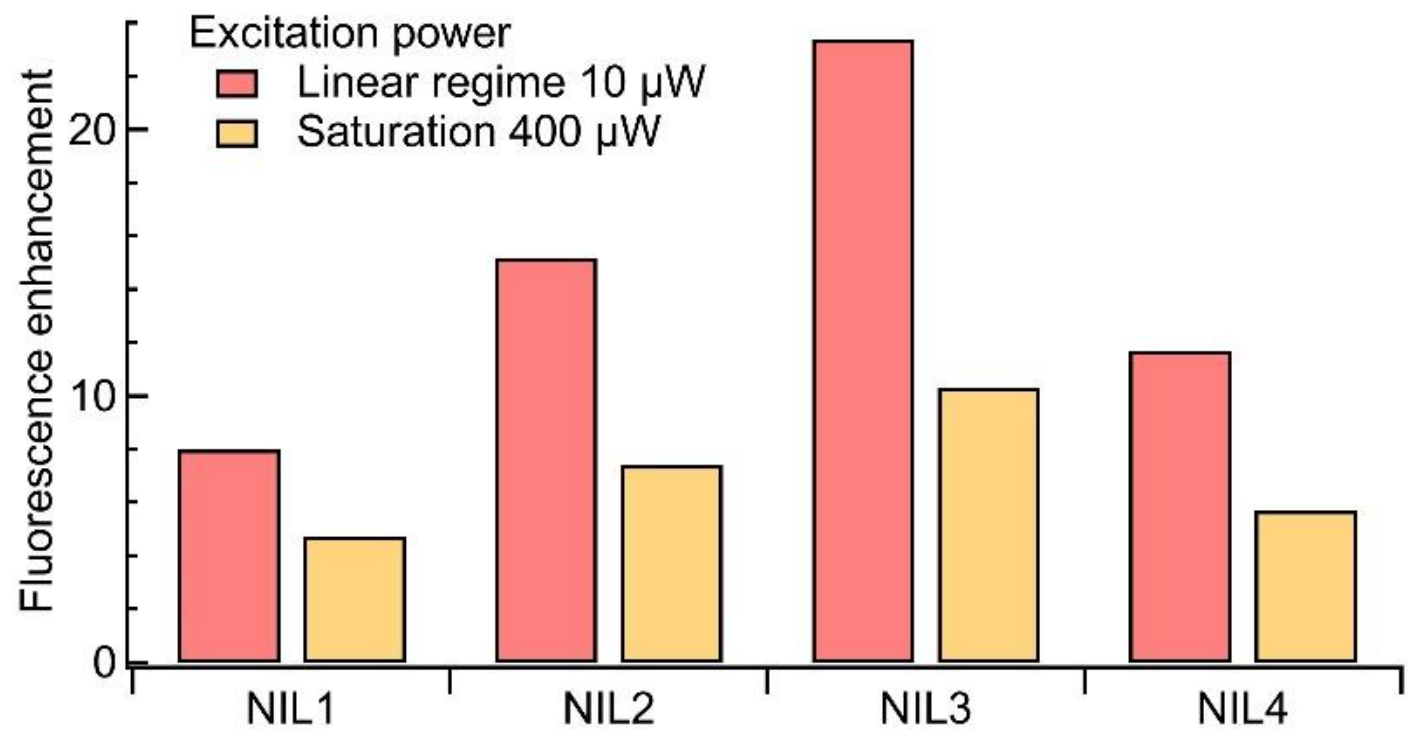

**Figure S5.** Enhancement factors of the fluorescence brightness per molecule respective to the confocal reference. The brightness data is presented in Fig. 3d. Due to saturation of the fluorescence process, the apparent enhancement factor is lower in the saturation regime (here 400 µW) than in the linear case (10 µW excitation power).

## S6. Background intensity as function of laser power

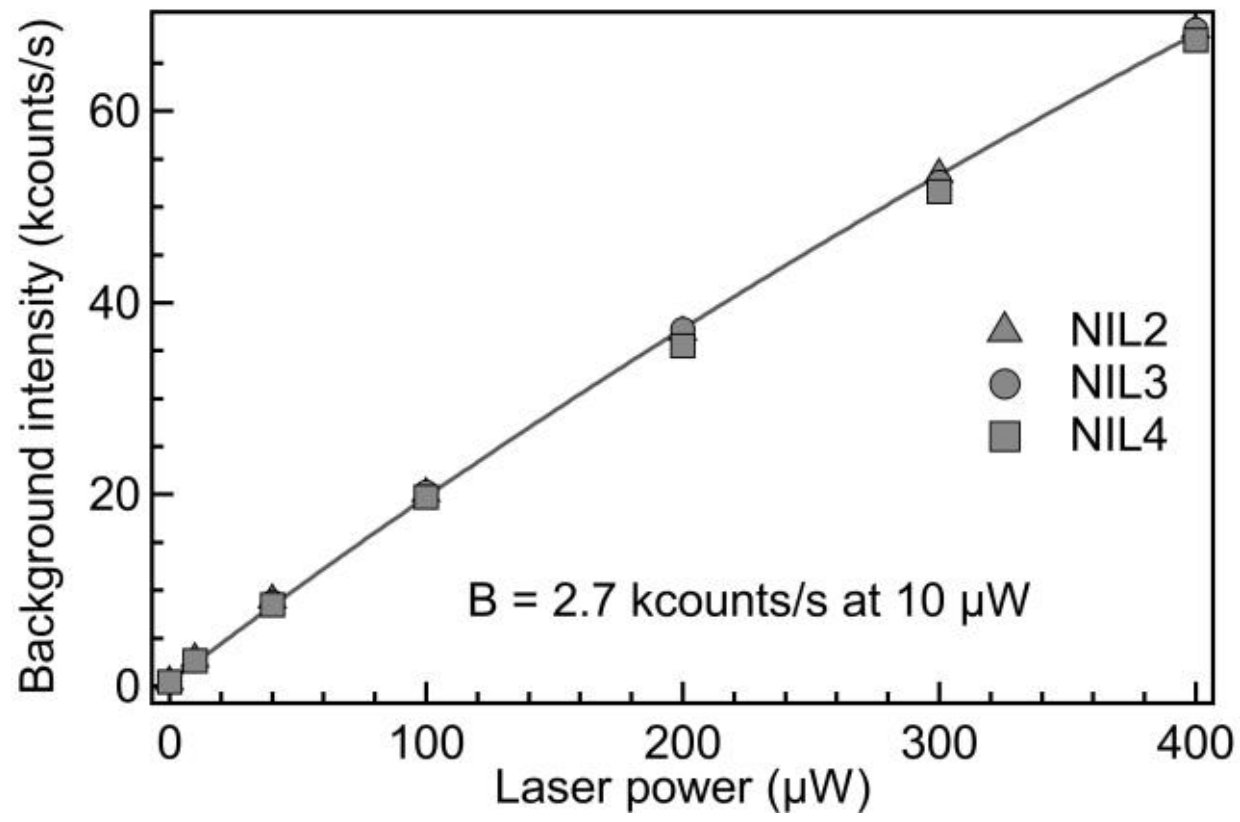

**Figure S6.** Background intensity dependence with the laser power, taken into account in the power dependence FCS analysis in Fig. 3d and S7. Most of the background stems from laser light back-reflected by the aluminum layer forming the ZMW. Additional filters could improve the situation, yet for this study we decided to work with a standard set of epifluorescence filters as described in the Methods section. At 10 µW excitation power used for most of this work, the background has a negligible influence.

## S7. Power dependence of the number of molecules and diffusion times

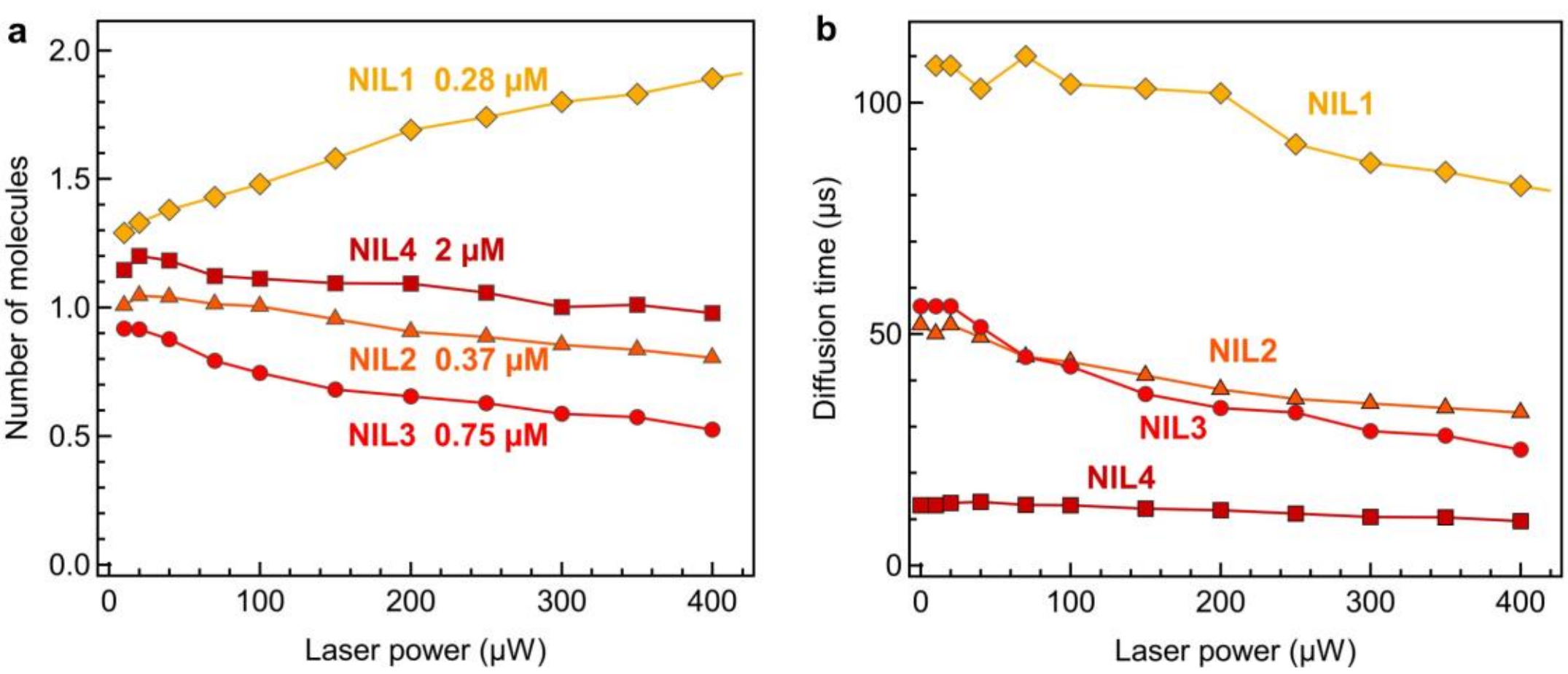

**Figure S7.** Number of molecules $N$ (a) and diffusion time (b) determined by FCS as function of the laser excitation power. These data correspond to the brightness experiment reported on Fig. 3d.

## S8. Scatter plots of detection volume, diffusion time and fluorescence enhancement

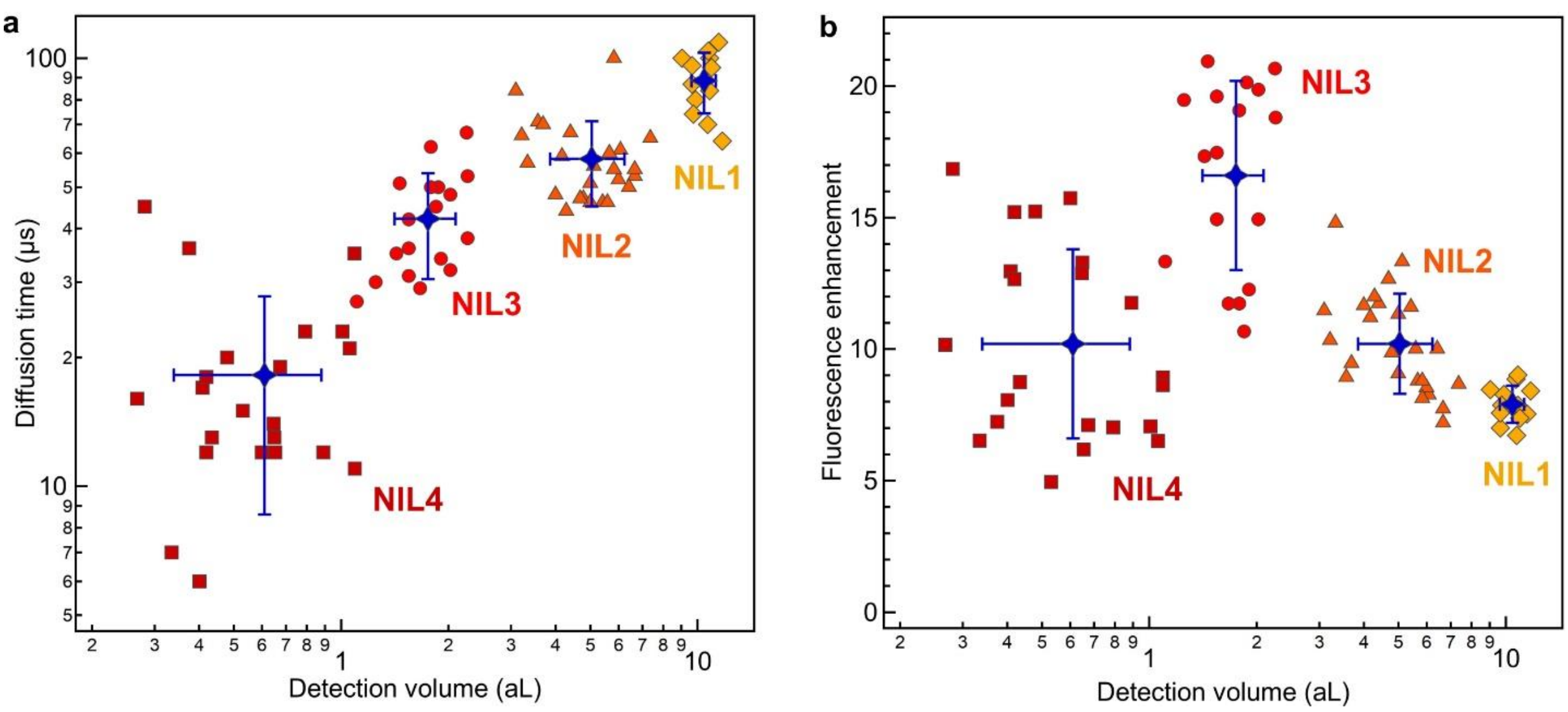

**Figure S8.** Diffusion time (a) and fluorescence brightness enhancement (b) as a function of the detection volume measured by FCS for each ZMW aperture. The blue crosses indicate the average value, with the error bars corresponding to one standard deviation. Each NIL*x* sample is represented by a distinct population.

## S9. Evolution of the relative errors across NIL samples

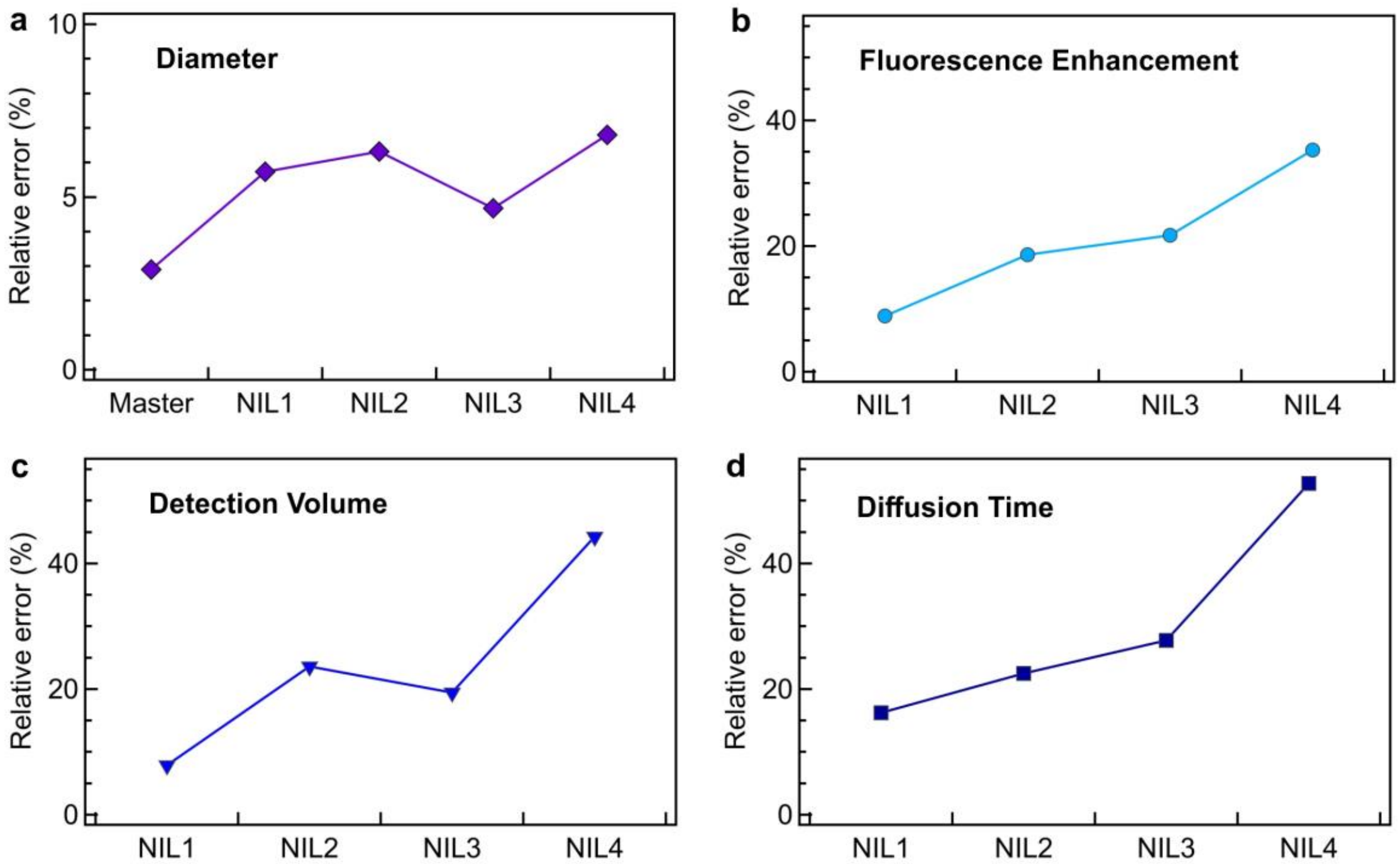

**Figure S9.** Relative errors defined as the ratio of the standard deviation by the average value for the ZMW diameter (a), the fluorescence brightness enhancement (b), the detection volume (c) and the FCS diffusion time (d). During the sequential NIL fabrication the deficiencies leading to statistical dispersion tend to accumulate. As a consequence, the relative errors increase with the NIL steps. In addition to this effect, the measured fluorescence quantities (brightness, volume, diffusion time) become more sensitive to small variations as the ZMW diameter is reduced and the electromagnetic field decays more strongly inside the ZMW.

## S10. Numerical simulations

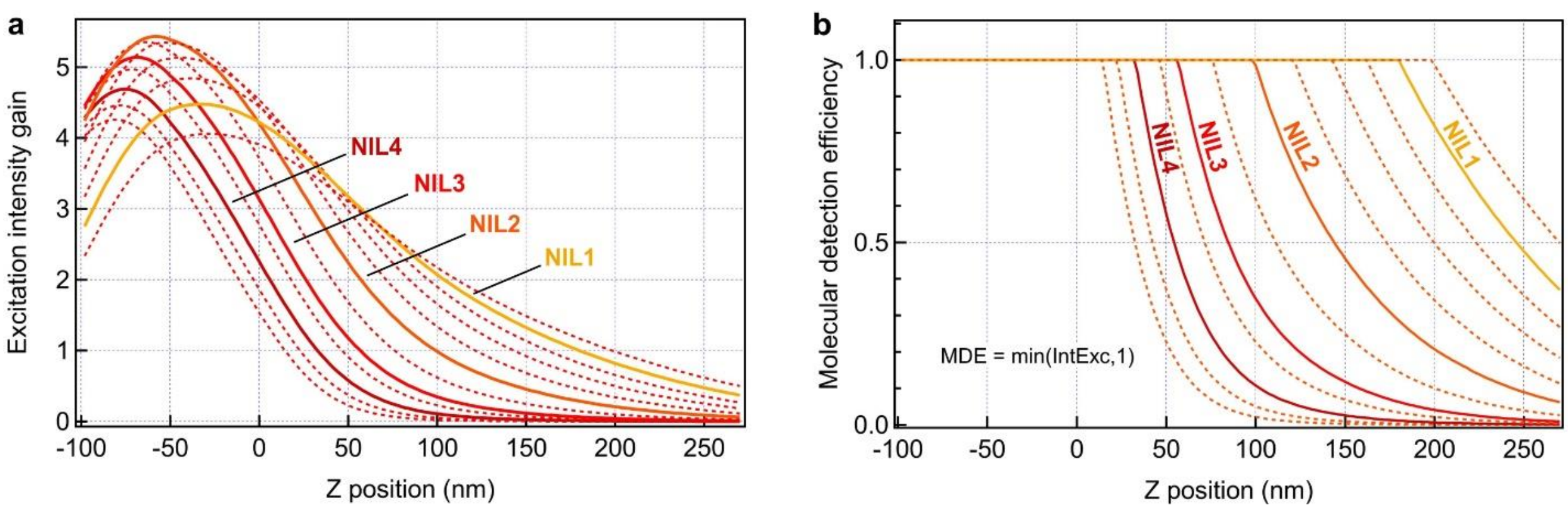

**Figure S10.** (a) Excitation intensity profiles $I_{ex}$ along the longitudinal ZMW dimension computed from the FDTD simulations in Fig. 4a. The Z=0 reference is taken at the glass-aluminum interface. The main diameters representative of the NILx samples used in this work are shown in thick lines. The dashed curves represent results for other ZMW diameters, with a diameter step of 14 nm between curves. (b) Molecular detection efficiency $MDE$ used to compute the detection volumes in Fig. 4b, defined as 1 if the excitation intensity at position Z exceeds 1, and taken equal to the excitation intensity otherwise.

## S11. Fifth-generation nanoimprint NIL5

5th generation NIL5

1 µm

500 nm

**Figure S11.** SEM images of the nanopillars corresponding to the 5th generation NIL5 fabricated using NIL4 as a master. The nanopillars display a conical shape. While the top diameter is 80 nm and corresponds well to the expected shrinkage reduction, the base is widened to 120 nm, leading to an apparent ZMW diameter similar to NIL4.

### S12. Comparison of NIL1 and NIL4 SEM images

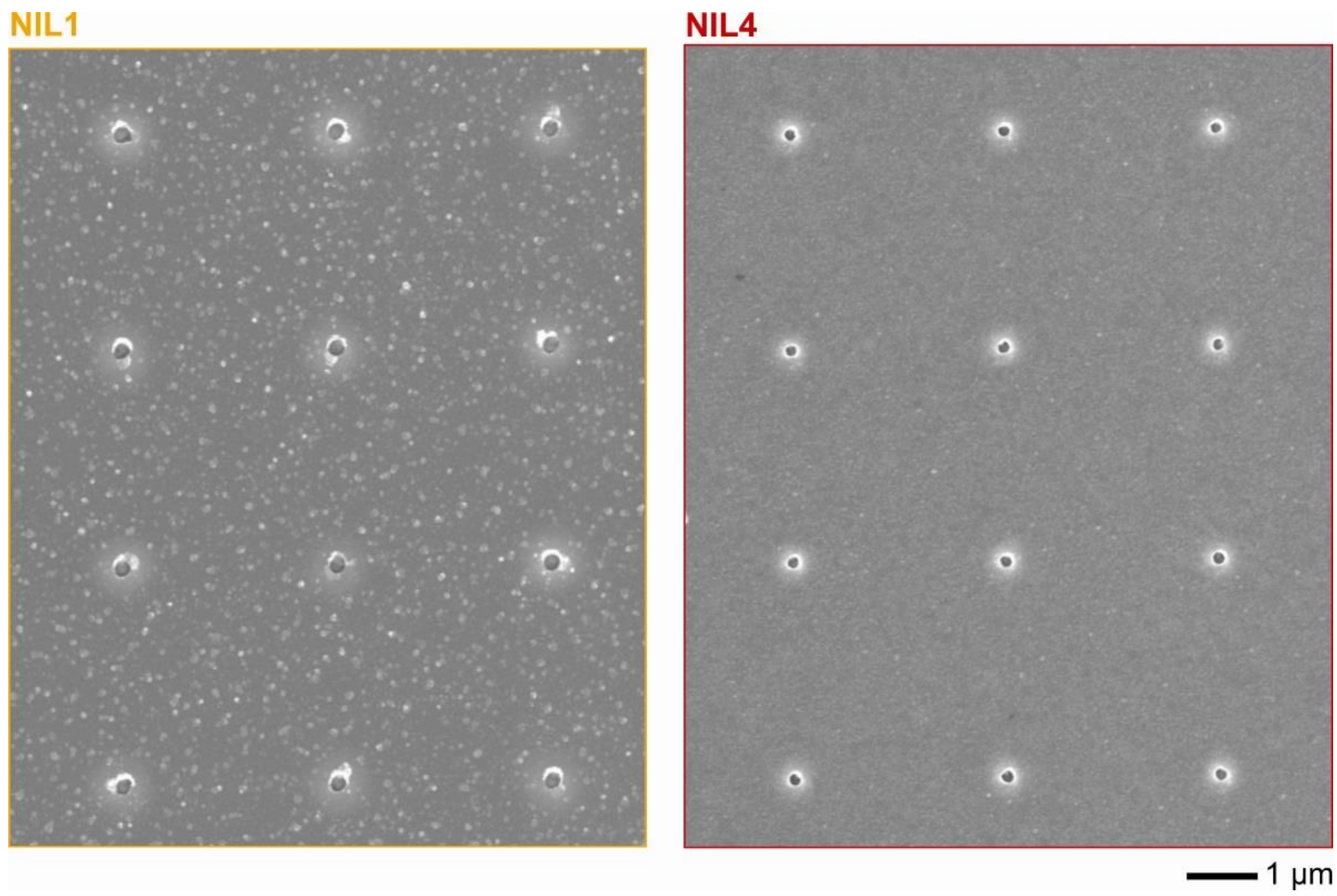

**Figure S12.** Comparison of SEM images of the ZMWs corresponding to the $1^{st}$ and $4^{th}$ generations NIL1 and NIL4, showing similar aperture shapes and preserved replication fidelity across generations. The minor shape and side-wall roughness variations present are primarily due to metal grains and remaining aluminum, rather than the sequential NIL process itself.

### S13. Size reduction across NIL generations

| | Diameter (nm) | Height (nm) | Shrinkage diameter (%) | Shrinkage height (%) |
|---|---|---|---|---|
| Master | 270 | 700 | n/a | n/a |
| NIL1 | 230 | 580 | 14,8 | 17,1 |
| NIL2 | 170 | 480 | 26,1 | 17,2 |
| NIL2 after RIE step | 170 | 600 | n/a | n/a |
| NIL3 | 140 | 460 | 17,6 | 23,3 |
| NIL4 | 115 | 390 | 17,9 | 15,2 |

**Table S1.** Summary of the characteristic dimensions of the nanopillars used in this work. The reduction in dimensions respective to the previous generation is expressed as the shrinkage percentage.